\documentclass[twocolumn]{aastex63}
\usepackage{color}
\usepackage[titletoc]{appendix}
\usepackage[fleqn]{amsmath}
\usepackage{amssymb}
\usepackage{mathtools}
\usepackage{float}
\usepackage{comment}
\usepackage{enumitem}
\usepackage{natbib}
\usepackage{graphicx}
\usepackage{bm}
\usepackage{totcount}

\shortauthors{Schultz et al.}
\shorttitle{OTS Mutual Inclinations}

\begin{document} 

\title{The distribution of mutual inclinations arising from the stellar quadrupole moment}

\author{Kathleen Schultz }
\affiliation{Department of Physics and Astronomy, University of Maine, Orono, ME 04469, USA}

\author{Christopher Spalding}
\altaffiliation{51 Pegasi b Fellow}
\affiliation{Department of Astrophysical Sciences, Princeton University, Princeton, NJ 08540, USA}
\affiliation{Department of Astronomy, Yale University, New Haven, CT 06511, USA}

\author{Konstantin Batygin}
\affiliation{Division of Geological and Planetary Sciences, Caltech, Pasadena, CA 91125, USA}

\email{kathleen.schultz@maine.edu}

\begin{abstract}
A large proportion of transiting planetary systems appear to possess only a single planet as opposed to multiple transiting planets. This excess of singles is indicative of significant mutual inclinations existing within a large number of planetary systems, but the origin of these misalignments is unclear. Moreover, recent observational characterization reveals that mutual inclinations tend to increase with proximity to the host star. These trends are both consistent with the dynamical influence of a strong quadrupolar potential arising from the host star during its early phase of rapid rotation, coupled with a non-zero stellar obliquity. Here, we simulate a population of planetary systems subject to the secular perturbation arising from a tilted, oblate host star as it contracts and spins down subsequent to planet formation. We demonstrate that this mechanism can reproduce the general increase in planet-planet mutual inclinations with proximity to the host star, and delineate a parameter space wherein the host star can drive dynamical instabilities. We suggest that approximately 5-10\% of low-mass Kepler systems are susceptible to this instability mechanism, suggesting that a significant number of single-transiting planets may truly be intrinsically single. We also report a novel connection between instability and stellar obliquity reduction and make predictions that can be tested within upcoming TESS observations.
\end{abstract}


\section{Introduction}

Our Solar system consists of 8 planets, orbiting the Sun within a few degrees of a common plane. This coplanar architecture stood as a primary motivation for the development of the ``nebular hypothesis" \citep{Kant1755,Laplace1796} -- the notion that planetary systems form in a low aspect-ratio disk of dust and gas. A natural expectation is that extrasolar planetary systems share this coplanar architecture. 

Individual extrasolar planetary inclinations are often difficult to measure \citep{fabrycky2014architecture,winn2015occurrence}. Nevertheless, at a population level, lower mutual inclinations typically result in a larger number of planets observed to transit any given star \citep{ragozzine2010value,lissauer2011architecture,he2019architectures}. To that end, observational surveys have repeatedly found an excess of systems containing only one transiting planet, 
relative to that which would be expected if extrasolar planetary systems typically possessed a Solar system-like, coplanar arrangement (known as the ``Kepler Dichotomy"; \citealt{johansen2012can,ballard2016kepler,zhu201830}). Henceforward in this work we will refer to systems with a lone transiting planet as ``single-transiting systems", and to the observed planet in question as a ``single-transiting planet", with similar terms for systems with multiple transiting planets. 

An excess of single-transiting systems suggests one of two physical scenarios  \citep{lissauer2011architecture,johansen2012can,ballard2016kepler}. First, perhaps these single-transiting planets are truly single, that is, they exist in systems lacking undiscovered non-transiting companions. Alternatively, the excess of singles might suggest that a substantial fraction of planetary systems possess multiple planets, but these planets' orbits are often inclined with one another by more than a few degrees \citep{johansen2012can,sanchis2014study, adams2020ultra}. To overcome this degeneracy, a combination of statistical analyses \citep{he2019architectures} and searches for signs of Transit-Timing Variations \citep{zhu201830} have been performed. These efforts imply that the excess of singles emerges predominantly by way of large mutual inclinations within higher-multiplicity systems. However, it is difficult to constrain the exact fraction of systems that are truly single, which have been suggested to constitute roughly 10\% of single-transiting planets \citep{he2020architecturesb}.

Given the expected coplanarity of protoplanetary disks, the widespread existence of mutual inclinations among planetary systems requires a physical explanation. Hypotheses involving planet-planet scattering require planets of excessively large mass \citep{johansen2012can}, and self-excitation appears insufficient \citep{becker2016oscillations}. The presence of an inclined exterior giant is capable of misaligning close-in planets \citep{hansen2017perturbation,becker2017effects,2017AJ....153...42L}, but requires that the giant obtained a mutual inclination in the first place \citep{pu2020strong}. While $\sim 30\%$ of close-in systems of super Earths possess exterior giants \citep{bryan2019excess}, their orbits tend to be aligned with inner groups of multiple transiting planets
\citep{masuda2020mutual}, but are poorly constrained around 
apparently-single transiting planets. 

An additional source of dynamical heating may arise from the host star itself \citep{spalding2016spin}. Specifically, while young, Sun-like stars spin relatively fast, acquiring a substantial quadrupole moment \citep{kraft1967studies,Ward1976secular}. In concert, observational characterization of the spin vectors of planet-hosting stars indicate that substantial values of stellar obliquity, the angle between the stellar spin pole and the planet orbit, are widespread \citep{winn2010hot,albrecht2012obliquities,winn2015occurrence,winn2017constraints}. Their origins remain uncertain, but viable pathways exist toward misalignments arising within the earliest 10-100\,Myr in many cases \citep{batygin2012primordial,spalding2014early,davies2019star}. Thus, a significant fraction of planetary systems likely experienced an early epoch during which they felt the gravitational influence of a rapidly-rotating, tilted star. 

The quadrupole moment of the central star tends to drive precession of the longitudes of ascending node among close-in planets \citep{murray1999solar}. Shorter-period planets precess faster, thereby misaligning the orbits of close-in planets within a given system owing to differential precession about the star's inclined spin pole. This mechanism has been shown to lead to misalignments \citep{spalding2016spin}, but also to dynamical instability for sufficiently-inclined and oblate stars \citep{spalding2018resilience}. 

A prediction of the oblate, tilted star (or ``OTS") mechanism is that planets residing closer to the host star are expected to exhibit larger mutual inclinations \citep{spalding2016spin}. Such a trend has emerged from recent characterization of multi-transiting planetary systems \citep{dai2018larger}. Specifically, planets residing closer than $\sim5$ stellar radii exhibit larger mutual inclinations than more distant orbits. Many of these planets fall into the category of Ultra-Short Period planets (USPs), which have orbital periods shorter than $\sim 1$\,day \citep{winn2018kepler}. Among this population, almost all are expected to possess exterior planets within 50\,days \citep{sanchis2014study,adams2020ultra}, further suggesting that the observed excess of single-transiting planets is primarily indicative of mutual inclinations. 

A direct application of the OTS mechanism, as described here, has successfully reproduced the observed mutual inclinations among USPs theoretically \citep{li2020mutual}. This work suggests that the trend observed by \citet{dai2018larger} is consistent with forcing from an oblate host star, at least in the case of static stellar oblateness. However, the stellar quadrupole moment falls substantially with time as the star spins down and contracts on the pre-main sequence \citep{bouvier2014angular}. The resulting sweeping of secular resonances is important for the excitation of mutual inclinations among the planetary orbits \citep{Ward1981solar,spalding2018resilience}. Moreover, if the star possesses obliquities beyond $\sim 30^\circ$, the stellar contraction may excite dynamical instabilities \citep{spalding2018resilience}. It is thus important to deduce the fraction of systems, including those hosting USPs, expected to undergo instabilities due to the perturbations from a contracting host star, which is the focus of the present study. 

In this work, we simulate a range of fabricated planetary systems. We inform our simulations using empirical constraints upon the typical masses and orbital separations within the Kepler dataset
\citep{millholland2017kepler, weiss2018california}. From these simulations, we obtain a relationship between the planetary orbital properties and the systems' eventual mutual inclinations. Consistently with \citet{li2020mutual}, we reproduce the trend seen in \citet{dai2018larger}, however, we identify a significant population of systems that undergo dynamical instability, leaving behind only a single planet. Though this population of intrinsic singles may be rarer, they provide additional observational tests of the OTS mechanism, which we describe. 

The remainder of this manuscript is organized as follows. In Section~\ref{sec: Methods}, we describe the set-up of our \textit{N}-body experiments, with the results presented in Section~\ref{sec: Results}. In Section~\ref{sec: Discussion} we discuss the requirements for and implications of instability, and assess the influence that various early-stage processes may have on the initial assumptions we have made in our model. We conclude in Section~\ref{sec: Conclusions} with predictions and notes for future work.

\section{Methods}\label{sec: Methods}
Our objective is to simulate the time evolution of the angle between two planets’ orbital planes in a variety of Kepler-like 2-planet systems. We suppose that initially, there exists a non-zero misalignment between the plane of the orbits and the stellar equator. The young host star possesses an appreciable quadrupole moment due to its own rapid rotation immediately following disk dispersal \citep{bouvier2014angular}. Over the subsequent hundreds of millions of years, the rotation rate slows owing to magnetic braking from stellar winds \citep{kraft1967studies}. To incorporate planet-planet interactions in addition to the decaying stellar quadrupole, we perform \textit{N}-body simulations using the software package \textit{MERCURY6} \citep{1999MNRAS.304..793C}, employing the hybrid Bulirsch-Stoer/symplectic algorithm. 

\subsection{Model set-up}

The gravitational potential of the central star was modeled up to quadrupole order \citep{murray1999solar}. Moreover, general relativistic apsidal precession was modelled by way of the addition of a dipole-like potential as described in \citet{nobili1986simulation} (see their equation 10). Once both of these effects are included within the simulation via a user-defined subroutine, the stellar potential in spherical coordinates is given by
\begin{align}\label{eq: potential}
V_\star(t)&=-\frac{G M_\star}{r}\Bigg[1+3\frac{GM_\star}{rc^2}\nonumber\\
&-\frac{3}{2}J_2(t)\bigg(\frac{R_\star}{r}\bigg)^2\mathcal{P}_2\big(\cos(\theta)\big)\Bigg],
\end{align}
where the second gravitational harmonic $J_2$ is explicitly time-dependent and $\mathcal{P}_2$ is the second degree Legendre polynomial. We define the stellar mass and radius as $M_\star$ and $R_\star$ while $G$ is Newton's gravitational constant and $c$ is the speed of light. The star's orientation is held fixed, and so $\theta$ is measured from the the stellar spin axis, while $r$ is the radial distance from the stellar center.

The magnitude of $J_2$ was forced to decay exponentially with time, from an initial value of $J_{2,0}$, over a timescale $\tau_\star$:
\begin{align}
J_{2}(t)=J_{2,0}\exp{\bigg(-\frac{t}{\tau_\star}\bigg)}.
\end{align}
While this expression may not exactly reflect the time-evolution of a star's $J_2$, the decay occurs on a timescale far exceeding the secular timescales of the orbits. Thus, the evolution is adiabatic and so the exact time-dependence of $J_2$ is unimportant \citep{henrard1982adiabatic,morbidelli2002modern}. In addition, as can be seen from Equation~\ref{eq: potential}, the dynamically-important quantity with respect to the stellar quadrupolar potential is not $J_2$ but $J_2 R_\star^2$. The stellar radius contracts by over a factor of 2 during the pre-main sequence, leading to a factor of 4 change in $J_2 R_\star^2$. For computational convenience, we encode all time-dependence within $J_2$ and hold $R_\star$ fixed at $R_\star=R_\odot$. This assumption slightly underestimates the quadrupolar potential at the earliest times. Nevertheless, Solar-type stars have undergone the majority of their contraction by the time the protoplanetary disk disperses \citep{gregory2016influence}, such that the effect of contraction should be of order unity.

The time step of our numerical simulations was set at 0.05 times the shortest planetary period, such that energy in the system at each step is conserved to within approximately $\sim10^{-6}$ of its initial value. 

\subsection{Initial conditions}

Our modelling seeks to deduce the importance of stellar obliquity and oblateness (parameterized through $J_2$) upon the orbital evolution of 2-planet systems. As mentioned above, we assume that the stellar spin-axis is held fixed, an assumption that stems from the much greater angular momentum of the star as compared with the planetary systems considered here \citep{spalding2018resilience}, as well as from the additional assumption that any distant objects excluded from the model exert negligible perturbations upon the star and system. For simplicity, we fix the star's spin axis as along the $z-$axis. Both planetary orbital planes are initially aligned with one another, and the orbits are circular. Stellar obliquity is defined as the angle between the host star's spin angular momentum vector and the angular momentum vector of the planets’ coplanar orbits; it is geometrically equivalent to the angle between those orbits and the host’s equatorial plane. Given our assumption of a $z-$axis-aligned stellar spin pole, the initial stellar obliquity is equivalent to the planetary orbital inclination angle\footnote{Note that stellar obliquity, in its most general case, is different for each orbit. Only by initializing the two planets to share the same orbital inclination, is it possible to define a single stellar obliquity in a 2-planet system.}. Thus, we parameterize stellar obliquity, represented as $\beta_\star$, in the form of nonzero initial orbital inclinations in our simulations.

Each system is constructed using a unique combination of stellar obliquity \(\beta_*\), initial stellar oblateness \(J_{2,0}\), the innermost planet’s semimajor axis \(a_1\), and the  planet mass \(m\). The range of \(a_1\) was selected to complement empirical data gathered by \cite{dai2018larger}. We choose to simulate 15 cases, given by
\begin{align}
    a_1/R_\star \in \{ 2, 3, 4, 5, 6, 7, 8, 9,
    10, 11, 12, 14, 16, 18, 20 \}.
\end{align}
These parameters are typical for the observed range of Kepler systems \citep{akeson2013nasa}. For simplicity, the host stars in our simulations are all of mass 1\(M_\odot\) and radius $1R_\odot=0.005$\,AU. Accordingly, the initial separations between host and nearest orbiting planet span between 0.01AU and 0.1AU.
 
Statistical evaluation of Kepler data has found that planets in 2-planet systems tend to be separated by about 20 mutual Hill radii on average \citep{weiss2018california}. The mutual Hill radius \(R_H\) between planets \textit{j} and\textit{ j+1} is defined as \citep{1993Icar..106..247G}
\begin{align}\label{hill}
 R_H=\bigg(\frac{m_j+m_{j+1}}{3M_\star}\bigg)^{1/3}\bigg(\frac{a_j+a_{j+1}}{2} \bigg)
\end{align}
for semi-major axes $a_j$, planetary masses $m_j$, and stellar mass \(M_*\). In our simulations, we fix planet-planet separation at 20 Hill radii \citep{weiss2018california}, such that
\begin{align}
  a_{j+1}-a_j=20R_H.
\end{align}
We parameterize our simulations in terms of $a_1$, the inner planet's semi-major axis, and set the two masses equal to one another. Thus, the value of $a_2$ is computed from $a_1$ as  
\begin{align}
a_2=a_1\bigg(\frac{1-10\mu^{1/3}}{1+10\mu^{1/3}}\bigg)
\end{align}
where we define $\mu\equiv 2m_1/3M_\star$. 

Within a given Kepler system, planets tend to be more similar to one another in both radius \citep{weiss2018california} and mass \citep{millholland2017kepler} than would be expected if they had been randomly assigned from the entire Kepler catalog. Critically for numerical simulations, this uniformity in mass allows us to select a single mass value \textit{m} and assign it to both planets in a simulated system. According to \cite{2014ApJ...783L...6W}, Keplerian sub-Neptunes have a mean mass of about 4.3 Earth masses; we have therefore selected planet masses of 1, 5, and 10\(M_\oplus\) in this work. However, we later find that planetary mass makes little difference to our conclusions. 

\begin{figure}
\centering
\includegraphics[trim=0cm 0cm 0cm 0cm, clip=true,width=1\columnwidth]{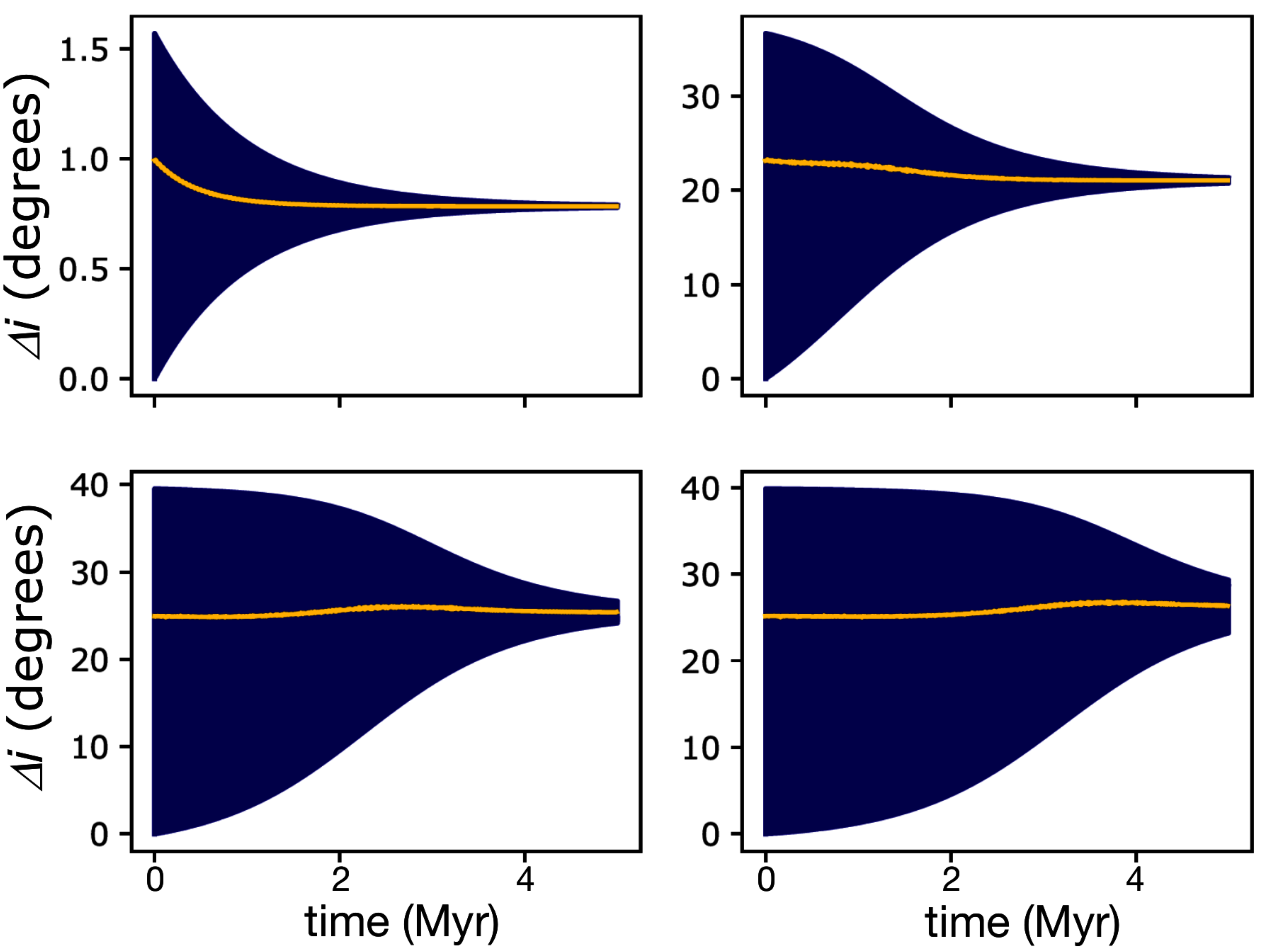}
\caption{Four examples of mutual inclination (\(\Delta i\)) evolution when two planets are integrated for 5 Myr around a tilted star with decaying oblateness, for different initial strengths of quadrupole moment. Each planet pair in this subset is identical, with \(a_1/R_*= 6\) and an average planet mass of 5\(M_\oplus\). Stellar obliquity is fixed at 20$^{\circ}$ in all four cases. From left to right, top to bottom, the systems were initialized with \(J_{2,0} \) values of \(1.98\times10^{-5}\), \(7.89\times10^{-4}\), \(3.14\times10^{-3}\), and \(7.89\times10^{-3}\), respectively. The gold curves denote mean \(\Delta i\) averaged across 10kyr intervals.}\label{4panel}
\end{figure}

A system is determined to have gone unstable when it loses one of its planets, either by ejection from the system or by collision with the host star. Previous investigations of the OTS mechanism have suggested that instability may occur in systems with stellar obliquities at or above about 30$^{\circ}$ \citep{spalding2016spin, spalding2018resilience}. Here, we seek to explore the expected properties both of systems that remain stable, and of those that undergo instability. Accordingly, we chose a distribution of obliquities spanning the ranges above and below the expected instability limit, with more resolution below $30^\circ$:
\begin{align}
 \beta_*\in\{ 1,6,10,20,30,50,70 \} ^\circ.
\end{align}
To first order in stellar spin rate $\Omega_\star$, the magnitude of $J_2$ at the beginning of integration may be related to the stellar Love number $k_2$, mass and radius through the approximate expression \citep{1939MNRAS..99..451S,Ward1976secular, spalding2016spin}
\begin{align}\label{J2Def}
J_{2,0}\approx\frac{1}{3}k_2\bigg(\frac{\Omega_\star^2}{GM_\star/R_\star^3}\bigg).
\end{align}
 The quantity $GM_\star/R_\star^3$ above may be identified as the square of the stellar break-up spin frequency.
 
 The Love number, \(k_2\), for a young star can vary by an order of magnitude depending upon whether the star is assumed as fully convective or fully radiative (e.g. \citealt{batygin2013magnetic}, \cite{1939MNRAS..99..451S}). At the beginning of our simulations, the star is yet to reach the main sequence, and is therefore likely to be fully convective, such that we assume an appropriate value of \(k_2=0.28\). Note that lower-mass stars leave the main sequence later than higher mass stars \citep{gregory2016influence}. Thus, $k_2$ may change from system to system, which constitutes an important area of follow-up in order to predict dependencies between mutual inclinations and stellar type. 
 
 Observations constrain the spin periods of young stars to between 1 and 10 days \citep{bouvier2014angular}. Given these empirical constraints, we created an initial set of \(J_{2,0}\) based on ten spin periods evenly distributed in log space within the $1-10$\,day range. We also selected an additional five \(J_{2,0}\) to explore the parameter space above $\sim 10^{-3}$; the proposed oblateness boundary between stability and instability in \citet{spalding2018resilience}. The entire set of initial \(J_{2,0}\) inputs in log space is

 \begin{align}
\log_{10}(J_{2,0})&\in\{-4.7, -4.5, -4.3, -4.1, -3.9,\nonumber\\
&-3.7, -3.5,-3.3, -3.1,-2.9,\nonumber\\
& -2.7,-2.5,-2.3, -2.1,-1.9\} 
\end{align}

The full set of simulations is thus comprised of 4725 planetary systems, each representing a unique combination of the four system parameters \(a_1\), \(m\), \(\beta_*\), and \(J_{2,0}\).
As described above, $J_2$ decays on a time scale $\tau_\star$. In reality, stars spin down over timescales of 0.1-1\,Gyr \citep{bouvier2014angular}. However, for the purposes of our simulations, we can adopt any value that's much longer than the planets’ forced precession frequencies \citep{henrard1982adiabatic,morbidelli2002modern}. We set the decay time scale to be $\tau_\star=1\,$Myr, and ran the simulations for 5\,Myr, or until only one planet remained. This simulation duration was sufficient to allow the stellar quadrupole moment to weaken such that mutual inclinations between planets in surviving systems settled into steady oscillations (see Figure~\ref{4panel}). A single mutual inclination (denoted by \(\Delta i\)) value is reported for each system that remained stable. We compute $\Delta i$ by averaging the mutual planet-planet inclinations over the last 10,000 years of the simulation.

\subsection{Empirical parameter distributions}

The data resulting from our suite of simulations provides a grainy, 4-dimensional grid that relates mutual inclination to $\beta_\star$, $J_{2,0}$, $m$ and $a_1$. Our goal is to use these relationships to construct a predicted distribution of mutual inclinations resulting from the observed set of these four parameters. To do this, we must first fit a continuous function that relates $\Delta i$ to \(J_{2,0}\) and \(\beta_*\) for any given $m$ and $a_1$. Next, we draw from empirically-informed distributions of $J_{2,0}$ and $\beta_\star$ in order to construct a predicted $\Delta i$ for each $a_1$ and $m$, marginalized over the stellar spin and tilt.

\begin{figure}
\centering
\includegraphics[trim=0cm 0cm 0cm 0cm, clip=true,width=1\columnwidth]{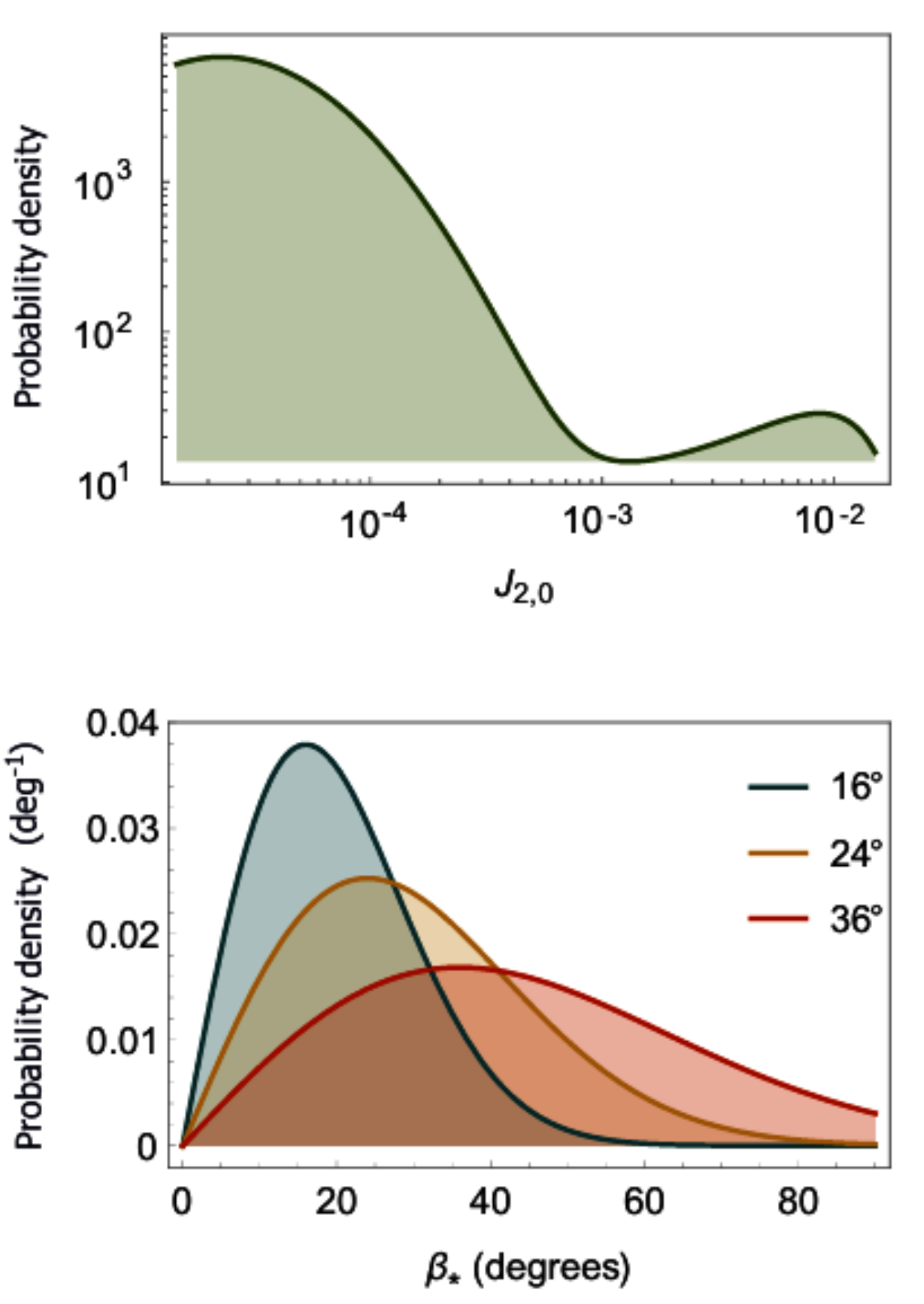}
\caption{Probability densities of distributions from which \(J_{2,0}\) and \(\beta_*\) were drawn when creating interpolation input sets. The curve in the upper panel is a fit to data collected by \cite{2016AJ....152..198K} and \cite{2005AJ....129..907B} from T Tauri stars in the Orion OB1 association. These sample stars possess masses within the range [0.47, 1.09] Solar masses and radii within [0.88, 1.66]\(R_*\). The Rayleigh curves in the lower panel were generated using scale parameters equal to 16 (mean stellar obliquity 20$^{\circ}$), 24 (mean 30$^{\circ}$), and 36 (mean 45$^{\circ}$).} \label{input_distribs}
\end{figure}

Both \(a_1\) and \(m\) were drawn from uniform distributions, on [0.01, 0.1] AU and [1,10] \(M_\oplus\), respectively. Observations of stellar mass and radius \citep{2005AJ....129..907B} and spin rate \citep{2016AJ....152..198K} in the Orion OB1 association allow us to estimate $J_2$ for a real population of young, rapidly rotating stars. Specifically, we compiled a list of stars that appear in both databases from \cite{2005AJ....129..907B} and \cite{2016AJ....152..198K}, then calculated each star’s \(J_2\) using Equation~\ref{J2Def}. The best fit distribution to the resulting \(J_2\) set is illustrated in the upper panel of Figure~\ref{input_distribs}. Our “artificial” \(J_{2,0}\) values were drawn from this distribution. Note the bimodal shape, likely indicative of the dichotomy between disk-hosting and disk-free stars \citep{rebull2018rotation}. Disk-hosting stars are thought to have their spin rates locked to the disk, whereas disk-free stars are free to contract and spin up \citep{armitage1996magnetic}.

The true distribution of evolved and early stellar obliquities remains an area of active research, in which a number of possible distributions have been proposed \citep{fabrycky2009exoplanetary,winn2017constraints}. To investigate the influence of different obliquity distributions on mutual inclination signatures in the OTS context, we choose four distributions from which to draw inputs: three Rayleigh distributions with means of 20$^{\circ}$, 30$^{\circ}$, and 45$^{\circ}$ (see bottom panel of Figure~\ref{input_distribs}),
and one uniform distribution on [0, 90]$^{\circ}$.
Consequently, we generated four separates sets of “artificial” \(\Delta i\) values, each based on randomly selected inputs from the three distributions of \(a_1\), \(m\), and \(J_{2,0}\) and one of the \(\beta_*\) distributions.

\section{Results}\label{sec: Results}
Figure~\ref{4panel} provides an example of the relationship between a stable planet pair’s \(\Delta i\) evolution and the strength of the central host’s quadrupole moment, for 20$^{\circ}$ stellar obliquity. In every case, the mean \(\Delta i\), represented by the gold curve, and maximum \(\Delta i\) come close to converging on a single value by the end of the simulation, which is expected when $J_2\rightarrow0$ and the two orbits precess around their total angular momentum vector. The system in the lower right panel was initiated with an unusually strong stellar quadrupole moment ($J_{2,0}\sim 8\times 10^{-3}$). During the first million years of this simulation, the mutual inclination oscillates between $\sim0^\circ$ and $2\beta_\star=40^\circ$, as expected in a system with a large, time-independent $J_2$ (see Equation~(16) of \citealt{spalding2016spin}). However, in the case considered here where $J_2$ decays with time, the maximum \(\Delta i\) also falls with time, reaching roughly $1.5\beta_\star$ by the end of the simulation. Thus, stellar spin down alone tends to slightly reduce mutual planet-planet inclinations when compared to their maximum, constant-spin value.

\begin{figure}
\centering
\includegraphics[trim=0cm 0cm 0cm 0cm, clip=true,width=1\columnwidth]{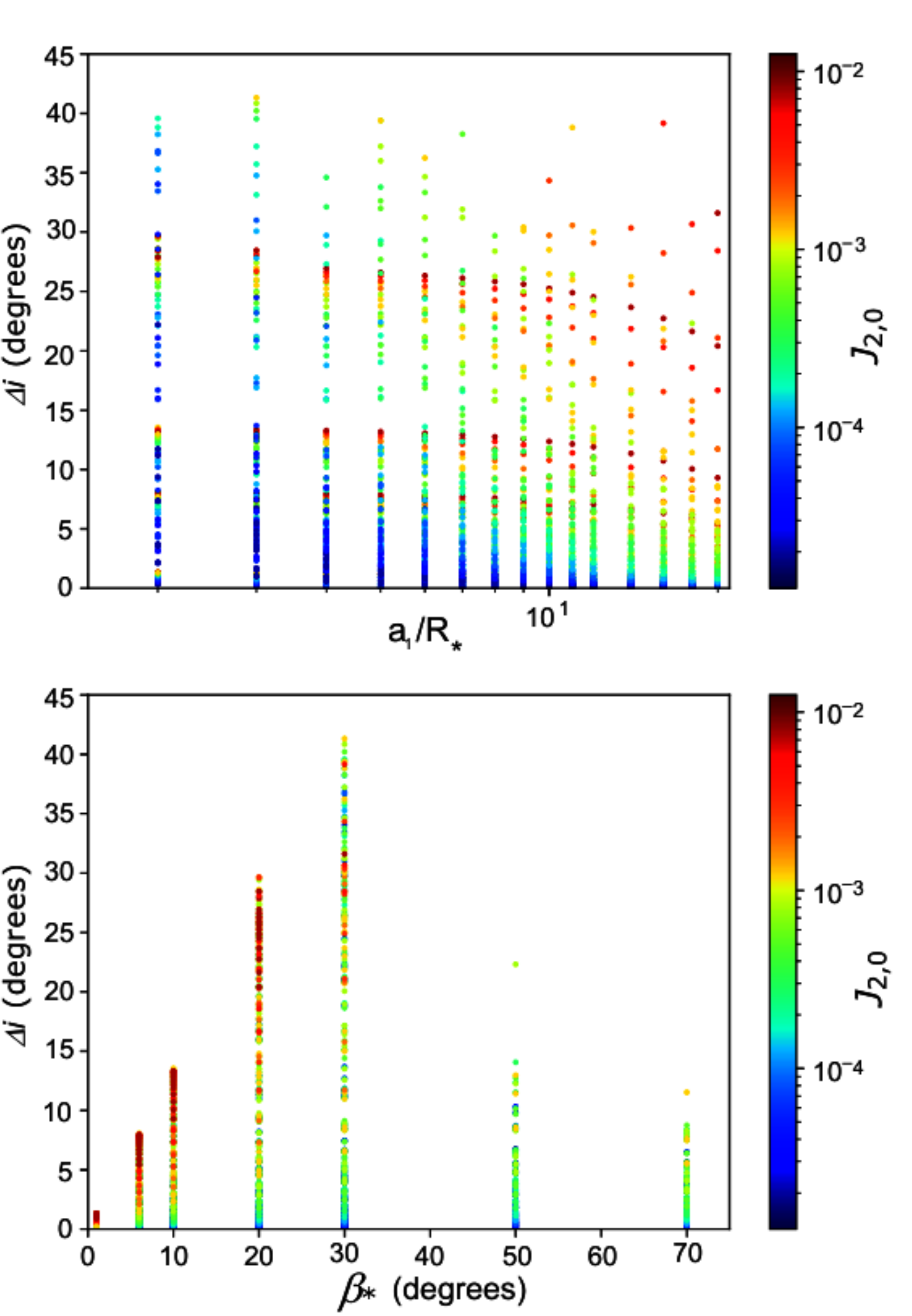}
\caption{Main integration results. Dot color in both panels corresponds to the value of \(J_{2,0} \) at which the represented system was initialized. Upper panel: Final mutual inclination between planets, as a function of separation between host star and innermost planet. One can consider this plot as being comprised of multiple 
layers of data, each associated with a different initial value of \(\beta_*\).
Dark red, high-oblateness bands appear at the maximum attainable values of \(\Delta i\) in each layer. The most prominent bands at ~14$^{\circ}$ and ~27$^{\circ}$ belong to the layers of data associated with 10$^{\circ}$ and 20$^{\circ}$ of stellar obliquity, respectively. Lower panel: Final mutual inclination as a function of stellar obliquity.}\label{rawdata}
\end{figure}

The mean final mutual inclinations of systems that remained intact at the end of integration are plotted in both panels of Figure~\ref{rawdata}. Note that mass is not a featured parameter in this figure, although it was a core variable in our initial inquiry. We found that any variation in planet mass within the range we adhered to was offset by the mass dependence of planet-planet separation (see Equation~\ref{hill}). Mutual inclinations were only marginally larger, and instability rates only slightly higher, for the lowest-mass systems over the highest-mass systems. As mass-dependent separation seems to be a feature of Kepler systems \citep{weiss2018california}, we conclude that the OTS mechanism is largely independent of planetary mass in Kepler-like, sub-giant planet pairs of equal mass.
\begin{figure}
\centering
\includegraphics[trim=0cm 0cm 0cm 0cm, clip=true,width=1\columnwidth]{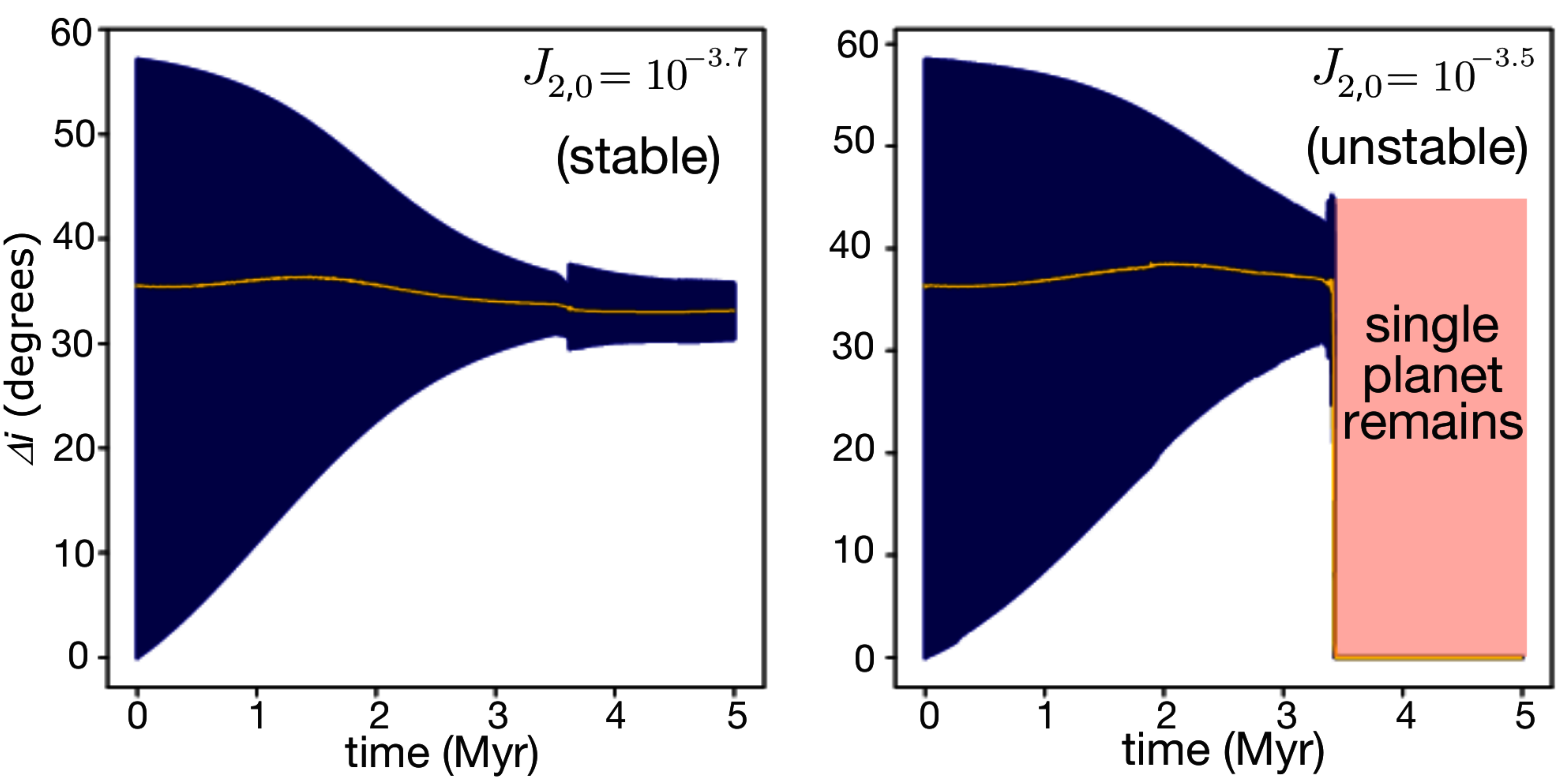}
\caption{Mutual inclination evolution of two planet pairs of identical mass, position, and stellar obliquity. The left-hand panel was initiated with \(J_{2,0}=10^{-3.7}\approx 2\times10^{-4}\), while the right-hand panel was initiated with \(J_{2,0}=10^{-3.5}\approx3.14\times10^{-4}\). As in Figure~\ref{4panel}, the gold curve represents average mutual inclination. Note that despite the relatively small increase in $J_{2,0}$, the right-hand system undergoes dynamical instability leaving an intrinsically-single system (red, shaded region). }\label{st_vs_unst}
\end{figure}

\begin{figure}
\centering
\includegraphics[trim=0cm 0cm 0cm 0cm, clip=true,width=1\columnwidth]{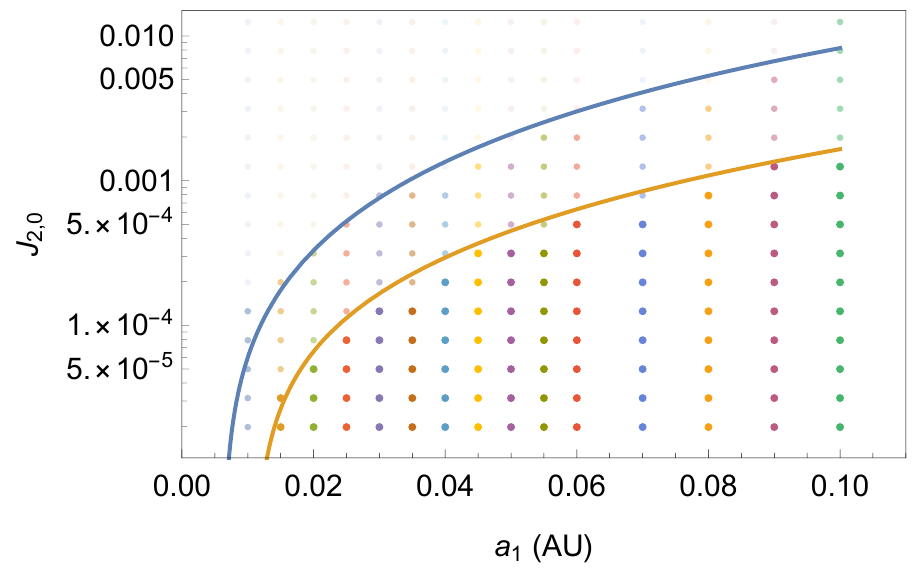}
\caption{The \(m=5M_\oplus\) subset of numerically integrated systems, plotted as initial oblateness versus innermost semimajor axis. More opaque points correspond to larger initial stellar obliquities. The blue curve marks the approximate values of \(J_{2,0}\) above which a system with 30$^{\circ}$ stellar obliquity will eventually go unstable. The orange curve denotes the same boundary for obliquities at or above about 50$^{\circ}$ (see equation~\ref{J2fits}).}\label{critJ2}
\end{figure}

It can be seen in both panels of Figure~\ref{rawdata} that, for a given \(\beta_*\), larger \(J_{2,0}\) leads to larger mean \(\Delta i\), as expected. Interestingly, no planet pairs reach mean mutual inclinations exceeding about 40$^{\circ}$. This is emphasized by the absence of a high-\(J_{2,0}\) band in the 30$^{\circ}$ “layer” of the upper panel, and the 30$^{\circ}$ column of the lower panel. This suggests that there exists a mean mutual inclination limit, between 35$^{\circ}$ and 40$^{\circ}$, above which planet pairs go unstable. Such a limit has been suggested to result from the crossing of a high-inclination secular resonance \citep{spalding2018resilience}. Our results here suggest that instability is not necessarily triggered when the planet pair crosses a critical, instantaneous \(\Delta i\), but rather when the time-averaged $\Delta i$ grows high enough.

To illustrate this, Figure~\ref{st_vs_unst} shows two alike systems’ \(\Delta i\) evolution, where one remained stable and the other did not. The doomed system goes unstable at approximately 36$^{\circ}$ \(\Delta i\) (3.3Myr), but not at the first instance of that value, and not at mutual inclinations larger than that value. The stable system peaks at the same value of \(\Delta i\), but doesn't demonstrate any signs of irregular excitation until much later in its evolution (about 3.6\,Myr, 34$^{\circ}$ \(\Delta i\)), after which it continues to evolve smoothly. Thus, the requirement for instability remains poorly constrained in a theoretical sense, but appears more closely tied to the system's average state than to its instantaneous state.

Our results allow us to more generally separate initial stellar conditions into those that favor stability and those that are hostile to it. In the bottom panel of Figure~\ref{rawdata}, for example, it is clear that planet pairs begin to go unstable for some obliquity between 20$^{\circ}$ and 30$^{\circ}$. Moreover a larger fraction go unstable at 50$^{\circ}$ than $30^\circ$, suggesting that the critical $J_{2,0}$ required for instability is dependent upon the stellar obliquity. It was previously estimated that the critical \(J_{2,0}\) separating stable from unstable dynamics lies close to $10^{-3}$ \citep{spalding2018resilience}, however, our results allow us to place more general constraints upon this requirement.

\begin{figure}
\centering
\includegraphics[trim=0cm 0cm 0cm 0cm, clip=true,width=1\columnwidth]{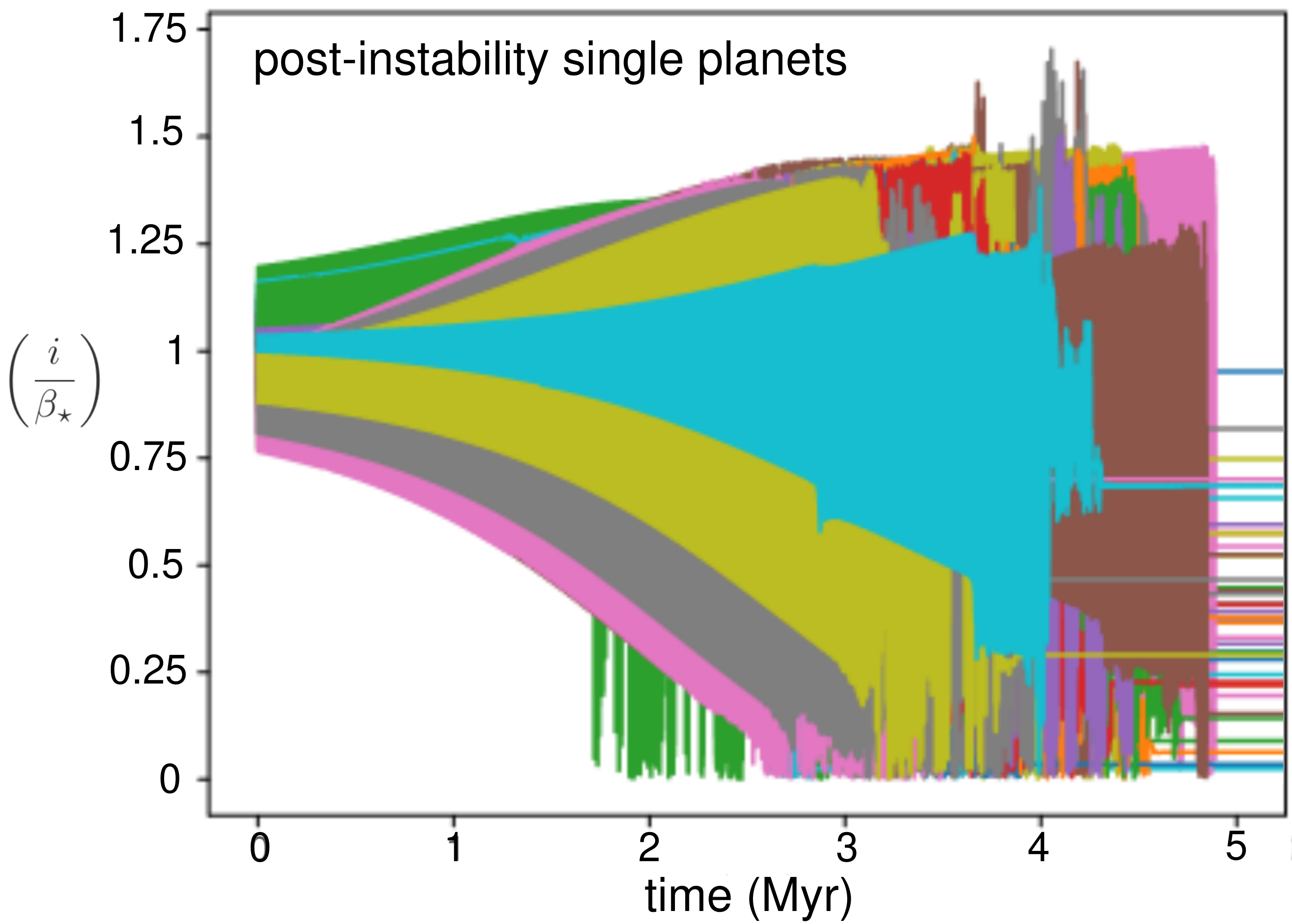}
\caption{Orbital inclination evolution of all singles that survived instability while in orbit around an initially 30$^{\circ}$-tilted star. Inclination has been scaled by stellar obliquity (see Figure~\ref{surv_dist}) The survivors’ final scaled inclinations have been extrapolated beyond the simulation’s 5Myr runtime to more clearly show their final values. }\label{surv_30}
\end{figure}

The critical $J_{2,0}$ required to initiate instability is illustrated in Figure~\ref{critJ2}. All of the surviving systems from each of the three ranges of obliquity (0-20$^{\circ}$, 30$^{\circ}$, and 50$^{\circ}$-70$^{\circ}$) are plotted according to \(J_{2,0}\) and \(a_1\). The points representing each range are plotted with increasing opacity, such that $0-20^\circ$ points are the faintest and $50^\circ$-70$^{\circ}$ the most visible. Points are not plotted in each case where instability occurred. Empirical fits to the stability limit for $30^\circ$ and $50^\circ$ have been overlaid, in order to approximate the boundary between stable and unstable \(J_{2,0}\) using an analytic function. The two curves of critical $J_2$ are represented by the functions

\begin{align}\label{J2fits}
J_{2,crit}= 
\begin{cases}
   -5\times 10^{-5}+0.003x+0.8x^2& \text{if } \beta_\star=30^\circ\\
   -5\times 10^{-5}+0.003x+0.14x^2,    & \text{if } \beta_\star\geq50^\circ
\end{cases}
\end{align}

where $x$ corresponds to the innermost semimajor axis $a_1$, in AU. The similarity between the above expressions to leading order in x suggests that, by tuning the coefficient of the quadratic term, one can predict the approximate value of \(J_{2,0}\) at which a close-in, similar-mass planet pair will go unstable for any stellar obliquity between 30 and 70 degrees. As our investigation did not include obliquities between 20$^{\circ}$ and 30$^{\circ}$, or above 70$^{\circ}$, we cannot extrapolate our instability predictions into these spaces. Nevertheless, with this data set, we have placed more precise constraints on the stellar properties required to eventually trigger instability in a given Kepler-like planet pair.

\begin{figure}
\centering
\includegraphics[trim=0cm 0cm 0cm 0cm, clip=true,width=1\columnwidth]{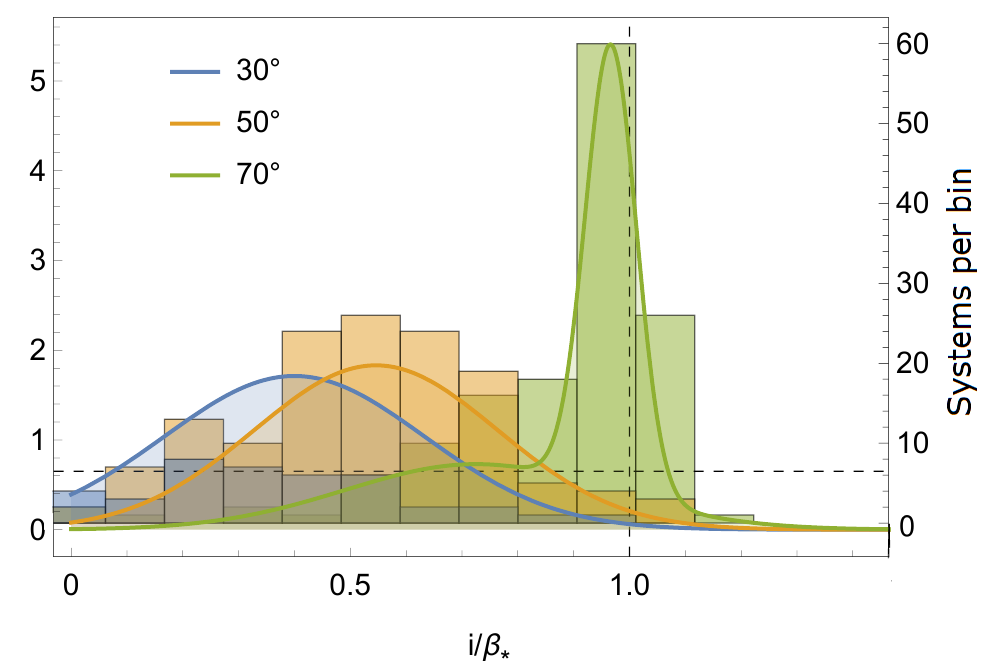}
\caption{Raw data (histograms) and probability
distributions of lone surviving planets' final orbital inclinations. Each inclination has been scaled, grouped, and colored according to the host's initial stellar obliquity. The horizontal dotted line approximates the form these distributions would take if they were entirely uniform (i.e. if post-instability orbital inclinations were entirely random).} \label{surv_dist}
\end{figure}

\begin{figure*}
\centering
\includegraphics[trim=0cm 0cm 0cm 0cm, clip=true,width=2\columnwidth]{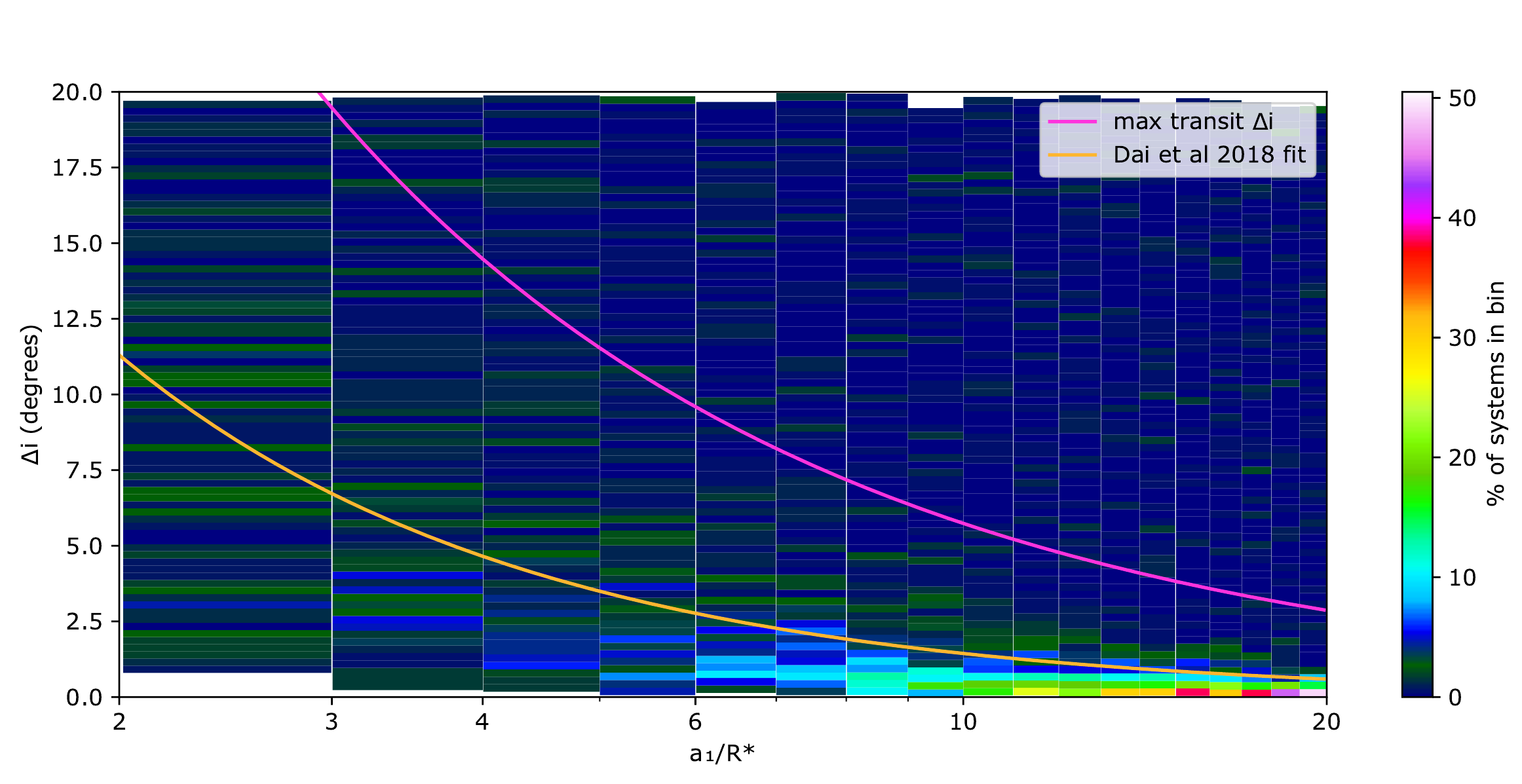}
\caption{A 2D histogram of interpolated \(\Delta i\) data based on our integration results, where the input \(\beta_\star\) set is Rayleigh-distributed with a mean of 20$^{\circ}$. Every column is a self-contained density plot: the color of a given bin expresses the number of systems in that bin as a percentage of the total number of systems in the column containing that bin. The overlaid curve in magenta marks the boundary between mutual inclinations that permit both planets to be observable via the transit method, and mutual inclinations that essentially restrict the pair to a single-transiting configuration. The orange curve
is the function found by \cite{dai2018larger} to best fit their observed mutual inclination data, converted from a series of distribution widths to a series of mean mutual inclinations for ease of comparison.}\label{column_dens}
\end{figure*}

The critical mutual inclination angle for instability here is similar to the critical inclination of $\sim 39^\circ$ required to enter the Kozai-Lidov resonance, in the case of a single planet perturbed by an exterior companion \citep{kozai1962secular}. This similarity was noted in \cite{spalding2018resilience}, but the critical angle $2\omega$ was not found to librate within resonance, counter to the case expected for the Kozai-Lidov resonance. Nevertheless, it remains intriguing that instability under the OTS mechanism shares traits with Kozai-Lidov resonance. The Kozai-Lidov mechanism has been studied mainly in systems where the perturbing and perturbed objects possess unlike masses and are widely separated \citep{naoz2011hot,naoz2016EKL}; it is unclear how the mechanism translates to systems comprised of closely-spaced, Earth-sized planets like those modelled here. Such an analysis is beyond the scope of this work.

A more surprising result of our simulations emerges upon examining the properties of systems after undergoing instability. Specifically, after instability, only a single planet survives. Naively, it might be expected that a single planet that results from instability should exhibit a larger spin-orbit misalignment than a system that remained stable. Figure~\ref{surv_30} presents the orbital inclination evolution of every single planet in the \(\beta_*\)=30$^{\circ}$ group that remained after instability, losing its partner in the process. Orbital inclination takes on the traditional Keplerian definition here and, in a single-planet system, is geometrically equivalent to stellar obliquity. The survivor’s final orbital inclination is expressed as a fraction of the host’s initial obliquity. Crucially, the surviving planet's final orbit was more aligned with the host’s equatorial plane than it was at the beginning of the simulation. 

The efficacy with which instability reduces stellar obliquity is further illustrated in Figure~\ref{surv_dist}. Here, scaled orbital inclinations of all survivors in the entire data set, grouped by initial stellar obliquity, are plotted in a histogram, along with probability densities for each group. These densities resemble normal distributions, with clear peaks and symmetry in the 30$^{\circ}$ and 50$^{\circ}$ cases. Moreover, the mean ratio of final orbital inclination to initial stellar obliquity increases with \(\beta_\star\), with a sudden shift in concentration and peak placement between the 50$^{\circ}$ and 70$^{\circ}$ populations. That is,  when a system with only moderate misalignment between host and planetary orbits goes unstable, the new stellar obliquity is fairly similar to its original value. This is in contrast with initially strongly misaligned systems, which experience more extreme obliquity erasure as a result of instability.

We elaborate upon this peculiar feature of our results later, but for now we suggest that stellar obliquity may not necessarily be expected to act as a signpost of dynamical instability. Rather, the process of instability tends to reduce stellar obliquities relative to systems where the two planetary orbits become misaligned, yet remain stable.

\subsection{Mutual inclinations versus orbital distance}

The mutual inclination of planet-pairs appears to increase close to the host star \citep{dai2018larger}, a pattern that is consistent with the OTS mechanism \citep{li2020mutual}. In this section, we replicate this result using our modeled population as outlined in section 2.3. 
We randomly sample from the uniform distributions of \(a_1\) and \(m\), and from the
distributions of $J_{2,0}$ and $\beta_\star$ illustrated in Figure~\ref{input_distribs}, choosing $20^\circ$ as the most common stellar obliquity  \citep{winn2017constraints}. We then input these four sampled values into an interpolation of our simulated data to produce an artificial mutual inclination value. 5000 total artificial \(\Delta i\) were generated this way and are plotted in Figure~\ref{column_dens}. The best-fit observed relationship found by \citet{dai2018larger} is overlaid in orange for comparison.

We note that the projected mutual inclinations between orbits as inferred in e.g., \citet{dai2018larger} serve as lower limits to the true mutual inclinations. This is true for two reasons. First, by requiring that the planets mutually transit, the inferred mutual inclinations are bounded above by $\sim R_\star/a$. Second, the orientations of the orbits with respect to the viewer affects the orientations and relative lengths of the transit chords. In the simplest case of 2 circular but mutually-inclined orbits, the mutual inclination as inferred from transit chord lengths is modulated by $\sim\sin(\theta)$ where $\theta$ is related to the viewing direction. If $\theta=0$, the intersection of the orbits is along the viewing direction and both chords are the same length. The average value of $\sin(\theta)$ from $0<\theta<\pi$ is $2/\pi$, and so in an approximate sense, our simulations over-estimate the  mutual inclinations by $\pi/2$. We ignore this effect given that $2/\pi$ is of order unity.

A qualitative inspection of Figure~\ref{column_dens} confirms that, when operated upon by the OTS mechanism, the closest-in planet pairs experience the greatest level of mutual inclination excitation. This result was arrived at using stellar properties drawn from observationally-motivated distributions, and thus arises naturally from the typical parameters characterizing planet-hosting stars. If the OTS mechanism is the main driver of excitation, the trend seen by \cite{dai2018larger}, and linked to the stellar quadrupole theoretically by \citet{li2020mutual}, will continue to strengthen as more close-in mutual inclinations are constrained.

Interestingly, our choice of the mean \(\beta_\star\) had minimal influence over the appearance of Figure~\ref{column_dens}. Thus, we cannot place significant constraints upon the true distribution of early stellar obliquities from these results alone. On the other hand, stellar obliquities of $\sim 20^\circ$ are well within observational constraints \citep{winn2017constraints}, hinting at a critical role played by the OTS mechanism in sculpting the final distribution of mutual inclinations.

\section{Discussion}\label{sec: Discussion}

All planetary systems begin their lives within a protoplanetary disk encircling a rapidly-rotating, inflated host star. While the gas disk is present, the planetary orbits are generally forced into coplanarity \citep{kley2012planet}. However, for $\sim 100$\,Myr subsequent to disk dispersal, the host star remains rapidly rotating \citep{bouvier2014angular,amard2016rotating}. If the host star possesses a non-zero obliquity, it is often capable of disrupting the primordial coplanarity of close-in systems. In contrast to these early stages of planetary system evolution, most planets are observed at ages exceeding a Gyr \citep{johnson2017california,petigura2017california}. By this time, their host stars have typically spun down to periods exceeding a week, leading to a highly reduced $J_2$. 

For the shortest period systems, the modern-day $J_2$ can still be dynamically significant \citep{li2020mutual,becker2020origin}, particularly for so-called ultra-short period planets (USPs), with periods shorter than 1\,day (corresponding to $a\lesssim 0.02\,AU=4R_\odot$). Nevertheless, these systems, too, must have passed through an earlier epoch of enhanced $J_2$, followed by decay of the $J_2$. Throughout this process, the dynamical system will inevitably sweep multiple secular resonances \citep{Ward1981solar} and potentially undergo instabilities \citep{spalding2018resilience}. Thus, it is important to determine the population-scale observational features expected to result from this ubiquitous process of stellar spin-down. 

In this work, we simulated a population of planetary systems subject to the dynamical influence of their host stars' quadrupole moment, taking account of the subsequent decay of their $J_2$ over time. Among systems that remained stable, shorter innermost orbital periods are associated with greater mutual inclinations (see Figure~\ref{column_dens}). This result was arrived at previously by \citet{li2020mutual}, and agrees with observational work \citep{dai2018larger}. However, our work indicates that when stellar obliquity exceeds $\sim 30^\circ$, large $J_2$ has the potential to drive dynamical instability. Moreover, we compute the the critical $J_2$ required to cause instability for various stellar obliquities (Figure~\ref{critJ2}).

Similarly to the excitation of mutual inclinations, close-in systems are more susceptible to instabilities (see Figure~\ref{critJ2}). This is consistent with a picture, described in \citet{spalding2018resilience}, wherein the stellar quadrupole drives mutual inclinations sufficiently high to access high-inclination secular resonances. It is passage through these resonances that eventually de-stabilizes the planetary system, a topic to which we now turn. 

\subsection{Instability and a mixed population}

The original puzzle of the ``Kepler Dichotomy" was that too many single-transiting systems exist to arise from a homogenous population of mutually-inclined multi-planet systems \citep{johansen2012can,ballard2016kepler,he2019architectures}. In general, this mystery can be solved by either a separate population of systems that intrinsically host only a single planet, or alternatively, by an additional population of multi-planet systems possessing larger typical mutual inclinations. Between these options, the latter is typically favoured by observations \citep{zhu201830,he2019architectures}. Moreover, the shortest period planets that are most susceptible to the stellar quadrupole (the USPs) are inferred to host exterior companions most of the time \citep{sanchis2014study,adams2020ultra}. Nevertheless, these studies have so far not ruled out that $\sim10\%$ of the signal from single-transiting planets may arise from intrinsically-single systems. Rather, a poorly-constrained, if sub-dominant fraction may be intrinsically single. 

In this work, using reasonable estimates for the planetary architectures and stellar parameters, we found that a substantial fraction of initially coplanar systems may undergo instability as a response to the oblate, tilted star. Such instability was previously revealed to act for a small number of specific systems \citep{spalding2018resilience}, and generally occurred only for stellar obliquities exceeding $\sim 30^\circ$ and $J_2\gtrsim 10^{-3}$. In this work, we likewise find that the onset of instability occurs near 30$^\circ$ of obliquity, but the precise tilt depends upon the system's $J_2$ and semi-major axis (see Figure~\ref{critJ2}). Across the entire parameter range, the value of $J_2$ required to destabilize a system varies by a factor of five.

In order to get a sense of the potential prevalence of instability from the OTS mechanism, suppose for definiteness that systems possessing both $\beta_\star>30^\circ$ and $J_{2,0}>10^{-3}$ undergo instability. If stellar obliquities follow a Rayleigh distribution peaked at $20^{\circ}$ \citep{winn2017constraints}, then about 32\% of stars would be tilted beyond $30^\circ$. In tandem, about 39\% of stars used to construct the upper panel of Figure~\ref{input_distribs} possess $J_{2,0}>10^{-3}$. Assuming that $\beta_\star$ and $J_{2,0}$ are statistically independent, these occurrences imply that about $13\%$ of systems like those studied here are susceptible to instability through the action of the stellar quadrupole potential. This number is only included as an estimate, given the uncertainties associated with stellar obliquities and initial spin rates. 

Moreover, a plethora of alternative pathways toward instability are available for close-in planetary systems \citep{chambers1996stability,ford2008origins,batygin2011instability,johansen2012can,petit2020path,pichierri2020onset,tamayo2020predicting}. Generally speaking, over $\sim 10\%$ of systems within the context of our simulations are expected to end up as true single planets. This fraction of intrinsic singles has been difficult to rule out in previous surveys \citep{sanchis2014study} and has been favored by the forward modelling approach of \citet{he2020architecturesb}. A promising avenue toward observational tests is to compare the obliquity distributions of single and multi-transiting systems using the line-of-sight rotational velocity of the host star \citep{morton2014obliquities,winn2017constraints}. 

\begin{figure}
\centering
\includegraphics[trim=0cm 0cm 0cm 0cm, clip=true,width=1\columnwidth]{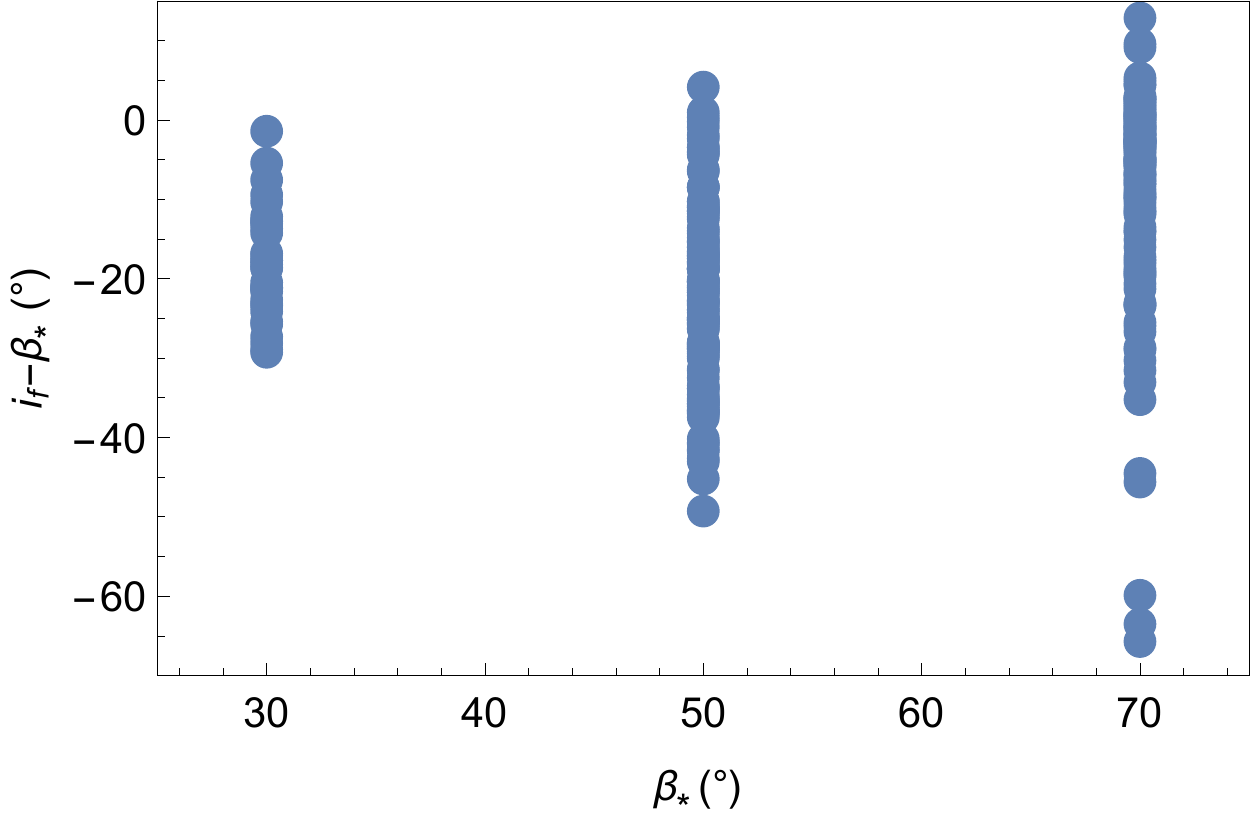}
\caption{
All surviving single-planet systems from our simulation, plotted according to the total orbital inclination shift (final minus initial) each single experienced as a result of instability. Negative values indicate relaxation from a high-inclination orbit onto a lower-inclination orbit, which is equivalent to a reduction in stellar obliquity.}\label{surv_change}
\end{figure}

Usually, it may be assumed that the singles will exhibit augmented spin-orbit misalignments if they arose from dynamical instability. However, our results suggest that when stellar oblateness drives instability, the one remaining transiting planet tends to possess a smaller spin-orbit misalignment than it did when it was younger (see Figures~\ref{surv_dist} \&~\ref{surv_change}). In other words, suppose that a population of primordially multi-planet systems emerge from the disk with a distribution of stellar obliquities $\beta_\star>30^\circ$. If all of these systems are driven to instability, then the mean stellar obliquity of the resulting transiting singles is smaller than that of the primordial population, including many examples with $\beta_\star<30^\circ$.

Observationally, these systems would appear to exhibit a low stellar obliquity, while providing no evidence of additional close-in planets. A possible hallmark of such ``violently aligned" systems would be high eccentricity, despite low inclination, such as K2-25b \citep{gaidos2020zodiacal,stefansson2020habitable}. Moreover, eccentricities of single-transiting planets appear higher than those of multis \citep{xie2016exoplanet,van2019orbital}, which indicates a violent history. It should be noted, however, that the remaining single is rarely brought to within $\sim10^\circ$ of the stellar equator, so would usually still arrive at its final orbit with a non-zero spin-orbit misalignment. Moreover, a central difficulty with using eccentricity as a tracer is that tides are expected to efficiently damp eccentricities in most systems residing close enough for the stellar quadrupole to be effective. Thus, eccentricity can only be used as a tracer of instability for a subset of single-transiting systems with longer tidal circularization times.

Broadening the scope beyond stellar oblateness-driven instability, we hypothesize that instability due to a misaligned external planetary perturber may result in an analogous degree of realignment between perturber and the planets surviving instability \citep{lai2017hiding,hansen2017perturbation}. Future efforts are required to further delineate the mechanics of post-stability realignment. 

\subsection{Migration}

Though we considered the time-evolution of the stellar $J_2$, we did not allow for migration among the planetary orbits. This is important because the orbits of USPs, analog to the closest-in planets in our simulations ($\sim0.01-0.02\,$AU), are closer to the star than the hypothesized inner edge of protoplanetary disks ($\sim 0.05\,$AU; \citealt{armitage1996magnetic,dullemond2010inner}). Moreover, the high temperatures existing interior to the disk typically lie above the sublimation limit for silicates \citep{flock2019planet}, preventing planet formation. Accordingly, lacking a local source of gas or dust, USPs are thought not to have formed \textit{in situ} but rather to have migrated inwards subsequent to disk-dispersal \citep{lee2017magnetospheric,winn2018kepler,petrovich2019ultra,millholland2020formation}.

If USPs migrate within the first $\sim 100$Myr after disk dispersal, they will likely experience the large values of $J_2$ assumed here. However, if migration occurs later, then the planets currently residing within $\sim 5$ stellar radii will have never been subjected to the enhanced $J_2$ of a star at their current $a_1$, but at a larger, primordial value of $a_1$. This would likely reduce instability rates and final mutual inclinations, with the result that far fewer of the close-in systems in Figure~\ref{column_dens} could achieve the large mutual inclinations observed by \cite{dai2018larger}. It is unclear whether the timescale on which inward tidal migration occurs is comparable to that of the weakening quadrupole moment.

In order to get a rough idea of when the $J_2$ is no longer sufficient, let us suppose that the perturbation of the inner planet from the outer member forces a nodal regression rate of \citep{murray1999solar}
\begin{align}
\nu_{pp}\approx - \frac{3 m_2}{4 M_\star}\bigg(\frac{a_1}{a_2}\bigg)^3 n_1.
\end{align}
The equation above is only correct to a factor of $\sim 2$ as we have assumed that $a_1\ll a_2$. However, for the purposes of the following calculation, no greater precision is required. In addition, the stellar oblateness forces a frequency of
\begin{align}\label{Stellar_Potential}
\nu_{\star}\approx - \frac{3}{2}J_2 \bigg(\frac{R_\star}{a_1}\bigg)^2 n_1.
\end{align}
If we use Equation~\ref{J2Def}, we may write the ratio of these two frequencies in terms of the stellar spin period, such that
\begin{align}
    &\frac{\nu_\star}{\nu_{pp}}\equiv \xi\approx\frac{8\pi^2}{3}k_2 \frac{a_2^3}{Gm_2}\bigg(\frac{R_\star}{a_1}\bigg)^5\frac{1}{P_\star^2}\nonumber\\
    &\sim 100\bigg(\frac{a_2}{0.06\textrm{au}}\bigg)^{3}\bigg(\frac{m_2}{6M_\oplus}\bigg)^{-1} \bigg(\frac{P_\star}{\textrm{day}}\bigg)^{-2} \bigg(\frac{a_1}{5R_\star}\bigg)^{-5}.
\end{align}

If the USP is migrating inwards over a timescale $\tau_a$ (such that $a_1\propto \exp(-t/\tau_a)$), but the star spins down over a timescale $\tau_{s}$ (such that $P_\star\propto \exp(-t/\tau_s)$), and we assume that $R_\star$ is fixed, then the ratio above evolves according to
\begin{align}
    \xi(t)\approx \xi_0 \exp\Bigg[-t\bigg(\frac{2}{\tau_s}-\frac{5 }{\tau_a}\bigg)\Bigg]
\end{align}
where $\xi_0\sim 100$, depending upon $a_2$ and $m_2$. Computing $d\xi/dt$, we obtain
\begin{align}
    \frac{1}{\xi}\frac{d\xi}{dt}=\frac{5\tau_s-2\tau_a}{\tau_a\tau_s},
\end{align}
which is positive (i.e., the star's influence increases with time) if $2\tau_a<5\tau_s$. In this case, the inner planet migrates inward rapidly enough to reach close proximity to the star before the quadrupole decays. In contrast, if $2\tau_a>5\tau_s$, the stellar quadrupole diminishes as the planet migrates inward, weakening the influence of the stellar oblateness (see also \citealt{becker2020origin}). 

The simple arguments above should be expanded upon in future work. Nevertheless, we see that the critical aspect of incorporating migration into the oblate tilted star framework lies in the relative timescales of migration and stellar spin-down. Both of these timescales are poorly constrained during the earliest 100\,Myrs of the system's evolution \citep{bouvier2014angular,spalding2019stellar}.

Initially, the spin-down timescales of Sun-like stars may range between $10-100\,$Myr \citep{garraffo2018revolution,spalding2019stellar}. However, over longer timescales, stars across a range of spectral types converge onto the so-called ``Skumanich" curve of $P_\star\propto t^{1/2}$ \citep{skumanich1972time,garraffo2018revolution}, with the spin-down timescale eventually reaching Gyr. The migration time of USPs remains an open question \citep{winn2018kepler,lee2017magnetospheric,pu2019low,petrovich2019ultra}, but it remains feasible that they migrate within the first 100\,Myr, thereby experiencing an early period with an enhanced stellar quadrupole.




\subsection{Initial conditions}

All of our simulations assumed that the planetary orbits began in a coplanar arrangement, but possessed a non-zero inclination relative to the stellar equator. This assumption is equivalent to declaring that the disk disperses instantaneously, thus preserving any primordial star-disk misalignment in the form of planet-star misalignment \citep{spalding2016spin}. However, in reality, disk dispersal is a poorly-understood process that may play out over timescales ranging from centuries to hundreds of thousands of years \citep{alexander2014dispersal}. If the disk disperses too slowly (that is, over several secular timescales of about $\sim10^3$ years), the inner planets will simply realign adiabatically with the stellar spin axis, unless a sufficiently massive, exterior companion planet is present \citep{spalding2020stellar}. 

Accordingly, our work here constitutes a thorough analysis of the consequences for planetary systems in the case where disk-dispersal occurs rapidly. In future work, it would be important to incorporate a range of disk-dispersal timescales into our framework, including the interaction between the planets and the disk material itself. However, these additions are beyond the scope of the current work. For now, we note that the importance of the OTS mechanism, and indeed that of multiple other early dynamical processes, hinges upon the specifics of disk dispersal. 

Finally, we mention that whereas we assumed that $J_2$ monotonically decreases with time, in reality stars that lose their disks are often still contracting onto the main sequence. When the disk is present, magnetic star-disk torques are hypothesized to enforce disk-locking, preventing the contracting star from spinning up \citep{armitage1996magnetic,rebull2018rotation}. Once the disk disperses, the star continues to contract, but approximately conserves angular momentum, forcing it to spin-up. This involves both a reduction in stellar radius and an increase in spin period, both of which are essential factors in determining the strength of the star’s quadrupole moment (see Equation~\ref{J2Def}) . If spin-up dominates physical contraction, we would see an initial upward trend in $J_2$, toward even larger values at the beginning of the post-disk phase. A $J_2$ spike could lead to more exaggerated final mutual inclinations and higher instability rates than are seen in our simulations. It is in the interest of future OTS inquiry that we question whether stellar contraction requires a change in the way we model time-dependent variation of the host’s quadrupole moment.

We may estimate the influence that pre-main sequence contraction has upon the stellar quadrupole moment using the following argument. First, we define the angular momentum of the host star as
\begin{align}
    \mathcal{J}_\star \equiv k M_\star \Omega_\star R_\star^2,
\end{align}
where $k$ is the dimensionless moment of inertia \citep{1939MNRAS..99..451S}. If $\mathcal{J}_\star$ is conserved ($d\mathcal{J}_\star=0$), and we neglect the small changes in $k$ and $M_\star$, we may relate the changes in $\Omega_\star$ to $R_\star$ with
\begin{align}\label{angular}
2\frac{dR_\star}{R_\star}=-\frac{d\Omega_\star}{\Omega_\star}.
\end{align}
Next, using Equations~\ref{J2Def} \&~\ref{Stellar_Potential}, we define the scaled quadrupole moment of the star as
\begin{align}
   \mathcal{L}\equiv J_2 R_\star^2\approx \frac{1}{3}k_2 \frac{\Omega_\star^2}{GM_\star}R_\star^5.
\end{align}
As the star contracts, the evolution of the quadrupole moment is given by
\begin{align}
   \frac{d\mathcal{L}}{\mathcal{L}}&= 2\frac{d\Omega_\star}{\Omega_\star}+5\frac{dR_\star}{R_\star}\nonumber\\
   &=-4\frac{dR_\star}{R_\star}+5\frac{dR_\star}{R_\star}\nonumber\\
   &=\frac{dR_\star}{R_\star}<0,
\end{align}
where the second line has used relationship~\ref{angular}. 

The simple calculation above indicates that, while the star is contracting immediately following disk dispersal, its quadrupole moment also tends to decay, despite the increasing stellar rotation rate. Thus, our approximation of a monotonically-decreasing $J_2$ subsequent to disk-dispersal accurately reflects the true evolution of PMS stars. 

We have not considered the evolution of $k_2$. In stars with $M_\star\gtrsim0.3 M_\odot$, a radiative core forms during the PMS phase, which reduces the value of $k_2$ by up to an order of magnitude \citep{batygin2013magnetic}. However, this effect will also tend to decrease the quadrupolar potential. Cumulatively, the stellar quadrupolar potential is at its highest early-on, and decreases from there. Throughout this decay, the system crosses secular resonances that may lead to instabilities \citep{Ward1981solar,spalding2018resilience}

\section{Conclusions}\label{sec: Conclusions}

The large number of single-transiting planets within the Kepler dataset, relative to multi-transiting systems, continues to elude a full explanation. While the majority of these singles are likely to possess unseen inclined companion planets, an uncertain fraction may be truly single. In this work, we provided a population-level exploration of the hypothesis that the host star's quadrupole moment excites mutual inclinations within primordially coplanar, close-in multi-planet systems. This oblate-tilted star (OTS) mechanism has previously been found effective in driving mutual inclinations among close-in planetary systems \citep{spalding2016spin,li2020mutual} and, if the star is tilted by more than $\sim 30^\circ$, also acts as a ubiquitous pathway toward instability \citep{spalding2018resilience}. However, before this work, the OTS mechanism had not been subjected to a population-scale analysis.

In this paper, we drew from recent observations in order to construct a suite of Kepler-like planet pairs. Using empirically-informed distributions of stellar obliquities and rotation periods, we simulated the orbital evolution of these systems in response to the host star's quadrupole moment immediately following disk-dispersal. Essential to the mechanism's effectiveness is the inclusion of early spin-down mechanics. Specifically, the stellar quadrupole decays due to spin-down and contraction of the star during the first $\sim100$ million years, causing the secular effect of the oblateness to sweep over multiple secular resonances \citep{Ward1981solar}. In contrast, a planet pair in orbit around a star with constant spin is unlikely to enter the secular resonance required to invoke instability \citep{spalding2018resilience}. Such a system will thus contribute only to the population of intrinsically 2-planet systems. 

Instability is the key to creating truly single planets in close orbits around their star. Broadly speaking, stellar spin periods below about 1-2\, days (or  $J_2\sim10^{-3}$), coupled with tilts exceeding $\sim30^\circ$ are capable of driving instability (Figure~\ref{critJ2}). Though stellar obliquities are poorly constrained for hosts of low-mass planets, up to 10\% of stars are expected to reside within this regime at some point during their histories. Accordingly the spin-down of the star and resulting incidence of instability must be considered when pursuing the question of the true nature of single-transiting planetary systems. We suggest that these singles are comprised of a mixture of lonely planets and those with excited, misaligned neighbors.

The exact proportion of apparent singles that are genuinely lonely depends on the rate at which stellar conditions – quadrupolar strength and obliquity – are favorable to large mutual inclination excitation and, ultimately, resonance. The 10\% instability rate we propose here is tentative, and will benefit from future investigations into early stellar misalignment and an improved understanding of the structure and rotation of young stars.

We have discovered a previously unreported phenomenon whereby the sole survivor of instability acquires a lower-inclination orbit than it possessed before the loss of its companion planet. That is, the spin-orbit misalignment of the one remaining planet is systematically lower than the spin-orbit misalignment of the initial 2-planet system (Figures~\ref{surv_dist} \& \ref{surv_change}). This effect tends to reduce the observable stellar obliquity to roughly half of the initial angle. Therefore, despite the fact that larger initial misalignments cause instability, the phenomenon of obliquity reduction may damp the expected association between instability and large stellar obliquities \citep{morton2014obliquities}. As modern stellar obliquity measurements continue to be an area of interest (e.g. \cite{winn2017constraints} and \cite{dai2020TKS}), we predict that the projected distribution of observed stellar obliquities in single-transit systems will be similar to that of multi-transit systems.

The outcome in a system undergoing evolution due to the OTS mechanism is sensitive to a number of early-stage factors that are not considered here. For example, this work did not include the brief interval of stellar spin-up that occurs while the star is still contracting onto the main sequence but the natal disk is removed. During this time, the cessation of star-disk torques facilitates a rapid decrease in rotational period and strength of quadrupolar moment. This “spin up” scenario widens the window in which inward-migrating planets can be excited via the OTS mechanism, and has yet to be incorporated into numerical OTS investigations.  

Further progress in the fields of disk dissipation, planet migration, young stellar modelling, and stellar obliquity evolution, particularly in understanding the relative time scales of each, is also critical and will improve our ability to draw conclusions about this early and dynamically rich part of a planetary system’s life cycle.

\section{Acknowledgements}
The authors would like to thank Neil Comins and Fei Dai for enlightening discussions. K.S. additionally thanks the Advanced Computing Group at the University of Maine for use of their facilities. C.S. is grateful for the generous support of the Heising-Simons Foundation.

\section{Data Availability}
The data underlying this article will be shared on reasonable request to the corresponding author.

\bibliographystyle{aasjournal}
\bibliography{main}

\end{document}